\documentclass[sigconf]{acmart}

\AtBeginDocument{%
  }

\setcopyright{acmlicensed}
\let\savedbaselinestretch\baselinestretch
\usepackage{listings}
\usepackage{multirow}
\usepackage{graphicx,adjustbox}
\usepackage{subcaption}
\usepackage{simplebnf}
\usepackage{threeparttable}
\usepackage{caption}
\usepackage{hyperref}
\usepackage{microtype}
\usepackage{booktabs}
\usepackage{tabularx}
\let\baselinestretch\savedbaselinestretch

\lstdefinelanguage{Kotlin}{
  morekeywords={val, var, fun, if, else, while, for, in, return, open, class, @Nullable},
  sensitive=true,
  morecomment=[l]{//},
  morecomment=[s]{/*}{*/},
  morestring=[b]",
  morestring=[s]{"""*}{*"""},
}

\newcommand{\toolName}{\textsf{CrossLangFuzzer}}
\newcommand{\kt}[1]{\href{https://youtrack.jetbrains.com/issue/KT-#1}{KT-#1}}
\newcommand{\scalaThree}[1]{\href{https://github.com/scala/scala3/issues/#1}{SCALA3-#1}}
\newcommand{\scalaTwo}[1]{\href{https://github.com/scala/bug/issues/#1}{SCALA2-#1}}
\newcommand{\groovy}[1]{\href{https://issues.apache.org/jira/browse/GROOVY-#1}{GROOVY-#1}}
\newcommand{\jdk}[1]{\href{https://bugs.openjdk.org/browse/JDK-#1?filter=allissues}{JDK-#1}}
\newcommand{\countKt}{15}
\newcommand{\countScalaTwo}{2}
\newcommand{\countScalaThree}{7}
\newcommand{\countGroovy}{4}
\newcommand{\countJava}{4}
\newcommand{\countAll}{32}

\title{\toolName{}: Differential Testing of Cross-Language JVM Compilers}

\author{Xiaotian Ma}
\orcid{0009-0008-9121-3656}
\affiliation{%
  \institution{Nanjing University of Science and Technology}
  \city{Nanjing}
  \country{China}
}
\email{xyzboom@njust.edu.cn}

\author{Qiong Feng}
\orcid{0000-0003-1667-8062}
\affiliation{%
  \institution{Nanjing University of Science and Technology}
  \city{Nanjing}
  \country{China}
}
\email{qiongfeng@njust.edu.cn}

\author{Yongqiang Tian}
\orcid{0000-0003-1644-2965}
\affiliation{%
  \institution{Monash University}
  \city{Melbourne}
  \country{Australia}
}
\email{yongqiang.tian@monash.edu}

\author{Wei Song}
\orcid{0000-0002-4324-3382}
\affiliation{%
  \institution{Nanjing University of Science and Technology}
  \city{Nanjing}
  \country{China}
}
\email{wsong@njust.edu.cn}

\author{Peng Liang}
\orcid{0000-0002-2056-5346}
\affiliation{%
  \institution{Wuhan University}
  \institution{School of Computer Science}
  \city{Wuhan}
  \country{China}
}
\email{liangp@whu.edu.cn}

\begin{document}

\begin{abstract}

Modern JVM software increasingly integrates multiple programming languages, such as Java, Kotlin, Groovy, and Scala, within a single application. Supporting such interoperability requires JVM compilers to perform cross-language compilation while reconciling subtle semantic differences across language boundaries. Errors in this process can lead to critical miscompilations, yet existing compiler testing techniques focus exclusively on isolated, single-language compilation.

To address this gap, we present \toolName{}, the first differential testing framework for cross-language JVM compilation. \toolName{} leverages the Kotlin compiler's unified intermediate representation (IR) to synthesize cross-language test programs. It further applies seven mutation operators to diversify generated test programs and improve bug-finding capability. Evaluated on the latest versions of five major JVM compilers, \toolName{} uncovered \countAll{} confirmed bugs, including \countKt{} in Kotlin, \countGroovy{} in Groovy, \countScalaThree{} in Scala 3, \countScalaTwo{} in Scala 2, and \countJava{} in Java.


\end{abstract}





\maketitle

\section{Introduction}\label{sec:background}

On the Java Virtual Machine (JVM), multiple programming languages—including Kotlin, Java, Groovy, and Scala—coexist and interoperate seamlessly~\cite{ardito:2020ist,mateus:2020esem}. Modern software systems increasingly exploit this multilingual ecosystem, routinely combining components written in different JVM languages within the same application~\cite{yang2024multi}. Supporting such interoperability requires JVM compilers to perform cross-language compilation: translating heterogeneous source programs into a common bytecode representation while correctly resolving dependencies and interactions across language boundaries. This process is inherently challenging. Unlike single-language compilation, cross-language compilation must reconcile semantic discrepancies among languages, including differences in type systems, nullability models, generics, variance rules, and method dispatch semantics. Subtle mistakes in handling these interactions can lead to miscompilations that are difficult to detect and diagnose.

Despite the growing importance of multilingual JVM software, existing compiler testing techniques predominantly target isolated single-language compilation pipelines~\cite{georgescu2024evolutionary, limpanukorn2025fuzzing, wang2025mut4all, chaliasos2021well, wang2025fuzzing, chen2020survey,csmith,YARPGen,ou2024mutators,DBLP:conf/ijcai/0001X0SC23,DBLP:conf/asplos/Wang00XS23}. Although several studies have explored testing language interoperability~\cite{dewey2014language,dewey2015fuzzing,chaliasos2022finding}, they do not systematically exercise the diverse semantic interactions arising in modern cross-language JVM applications. Consequently, current approaches lack the multi-language context necessary to expose faults at language boundaries, leaving a substantial class of cross-language compilation bugs largely unexplored.

To address this gap, we propose \textbf{\toolName{}}, a differential testing framework for cross-language JVM compilation. \toolName{} constructs semantically rich test programs using the Kotlin compiler's internal intermediate representation (IR), which provides a unified representation of program semantics independent of any specific JVM language\cite{rong2024irfuzzer, ates2026mlir}. The generated IR programs are subsequently rendered into Kotlin, Java, Groovy, or Scala source code, enabling the systematic creation of multilingual programs that exercise interactions across language boundaries. To further improve test diversity, \toolName{} applies seven mutation operators to the generated programs after the initial synthesis phase. By differentially testing multiple versions of JVM compilers, \toolName{} has uncovered \countAll{} confirmed bugs in the latest versions of five major JVM compilers. 

\section{Research Motivation}
\label{sec:motivation}

Although Kotlin is designed to interoperate seamlessly with Java, there are syntactic and semantic differences between Kotlin and Java (e.g., the latest Java versions still use raw types for backward compatibility with pre-generics Java code, while Kotlin does not support raw types). These differences pose challenges for Kotlin and Java compilers when processing cross-language programs.

\begin{figure}[!htbp]
    \centering
    \begin{minipage}[c]{\textwidth}
        \begin{lstlisting}[language=Java]
// Java Files
public interface A<T> {
    void foo(List<T> list);
}
public abstract class B implements A<String> {
    @Override
    public final void foo(List list) {}
}
public class C extends B implements A<String> {}
// Kotlin File
class X: C() //compiler error
        \end{lstlisting}
    \end{minipage}
    \caption{\kt{55822}: The Kotlin compiler rejects a valid Java program with raw type arguments}
    \label{fig:example1}
\end{figure}

Figure~\ref{fig:example1} illustrates a bug in the Kotlin compiler when handling raw types from Java. In Class \texttt{B}, the method \texttt{foo} overrides \texttt{A.foo} using a raw type. Class \texttt{C} extends Class \texttt{B} and implements interface \texttt{A} with the generic argument \texttt{String}. When a Kotlin Class \texttt{X} inherits from Class \texttt{C}, the Kotlin compiler rejects this valid program, reporting: ``Class `\texttt{X}' is not abstract and does not implement abstract base class member \texttt{public abstract fun foo(list: (Mutable)List<String!>!): Unit}''. If Class \texttt{X} were a Java class, the Java compiler would correctly accept this code.

In this example, the Kotlin compiler fails to recognize \texttt{B.foo} as an override of \texttt{A.foo} because it treats the raw type \texttt{List} as a wildcard type \texttt{List<*>} (for the reader's understanding, this is a simplified explanation; in actual Kotlin-Java interoperability, the raw type \texttt{List} is treated by Kotlin as a flexible type \texttt{List<*>..MutableList<*>?}). Since the wildcard \texttt{*} type does not match the generic argument \texttt{String} in interface \texttt{A<String>}, the compiler does not recognize \texttt{B.foo} as an override of \texttt{A.foo}. The Kotlin development team has previously located and fixed this issue (\kt{55822}); we use it here purely as an illustrative example of the bug class, and it is not among the \countAll{} bugs newly reported in this work. 

This case demonstrates a Kotlin compiler bug caused by improper handling of the inconsistency between Kotlin and Java raw types. Beyond raw types, Kotlin, Scala, Groovy, and Java also differ in inheritance and generics design, null handling, and declaration-site variance~\cite{kotlinSpec}. Identifying such cross-language compilation issues is crucial for assessing the robustness of Kotlin/Java compilers. Therefore, establishing systematic cross-language compilation testing is essential for improving the interoperability and reliability of the JVM ecosystem.

\section{Design and Implementation}\label{sec:design}

This section details the architecture of \toolName{}, its unified intermediate representation (IR), and its mutation strategies. \toolName{} extends our prior prototype~\cite{feng2025finding} by refactoring and enriching the IR, introducing additional mutations, and adding automated test-case reduction. These extensions enabled \toolName{} to discover 8 additional bugs, raising the total from 24 to 32. We also package the tool as Docker images for ease of use.

\subsection{Overall Framework}\label{subsec:framework}

As illustrated in Figure~\ref{fig:framework}, \toolName{} executes an automated pipeline spanning four major components:

\begin{figure}[!htbp]
    \centering
    \includegraphics[width=0.99\linewidth]{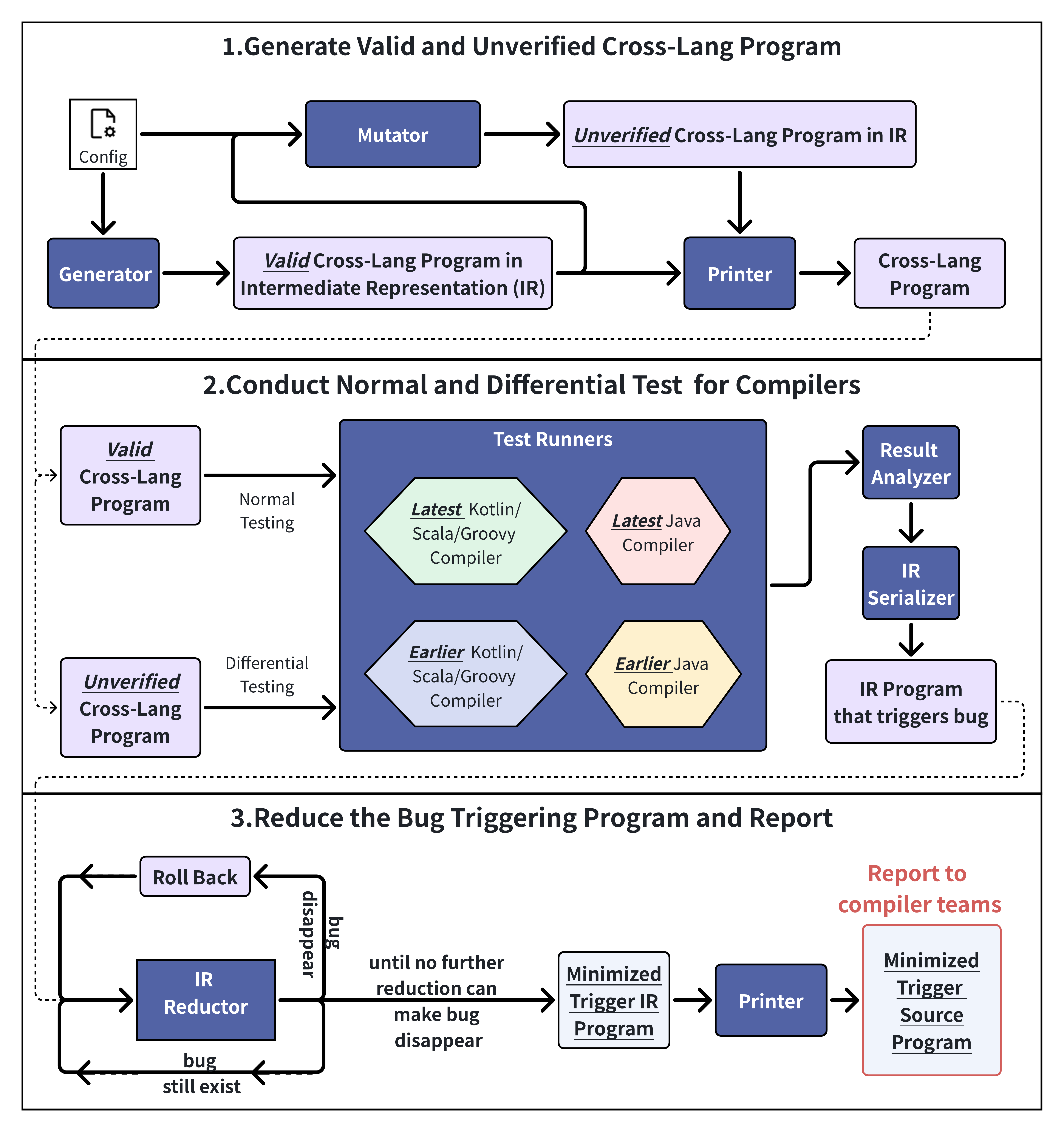}
    \caption{\toolName{} Overall Framework}
    \label{fig:framework}
\end{figure}

\begin{table*}[t]
    \centering
    \caption{Mutation Operators in \toolName{}}
    \label{tbl:mutations}
    \begin{tabular}{lp{0.55\linewidth}}
        \toprule
        \textbf{Mutation Operator} & \textbf{Description} \\
        \midrule
        \texttt{mutateGenericArgumentInParent} & Replaces a type argument within a parent class or interface. \\
                                               & \textit{Example:} \texttt{List<String>} $\rightarrow$ \texttt{List<Int>} \\
        \addlinespace
        \texttt{removeOverrideMemberFunction} & Strips an overridden method body, transforming it into an empty stub while preserving its signature. \\
        \addlinespace
        \texttt{mutateGenericArgumentInMemberFunctionParameter} & Replaces a nested type argument inside a function parameter. \\
                                                                & \textit{Example:} \texttt{Map<K, V>} $\rightarrow$ \texttt{Map<Int, String>} \\
        \addlinespace
        \texttt{mutateParameterNullability} & Toggles the nullability modifier of a targeted function parameter. \\
                                            & \textit{Example:} \texttt{x: String} $\rightarrow$ \texttt{x: String?} \\
        \addlinespace
        \texttt{mutateClassTypeParameterUpperBoundNullability} & Toggles whether a class type parameter's upper bound allows null values. \\
                                                               & \textit{Example:} \texttt{T <: Any} $\rightarrow$ \texttt{T <: Any?} \\
        \addlinespace
        \texttt{mutateClassTypeParameterUpperBound} & Replaces the upper bound of a class type parameter with an entirely distinct type. \\
                                                   & \textit{Example:} \texttt{T <: Any} $\rightarrow$ \texttt{T <: Comparable<T>} \\
                                                   \addlinespace
        \texttt{shuffleLanguage} & Reassigns the target language of IR classes. \\
                                                   & \textit{Example:} \texttt{(Kotlin,Java)} $\rightarrow$ \texttt{(Scala,Java)}\\
        \bottomrule
    \end{tabular}
\end{table*}

    (1) \textbf{Generator:} Given a configuration file which specifies target JVM languages, the generator synthesizes an initial cross-language program modeled in our custom IR. The generator is \textit{valid by construction}: integrated semantic constraint-checking guarantees that the generated class hierarchies and type assignments strictly conform to JVM inheritance rules.
    
    (2) \textbf{Mutator:} To enhance test-case diversity, \toolName{} passes the initial valid IR to a mutation engine, which deliberately applies structural modifications without enforcing semantic validation. This lack of validation concerns semantic well-typedness, not syntactic validity. For programs taken directly from the generator without mutation—which are well-typed by construction—any divergence in accept/reject behavior across compilers is flagged as a candidate bug. For programs altered by mutation, which may be ill-typed, we likewise treat accept-versus-reject divergence as the trigger, but only as a candidate: because compilers may legitimately differ on invalid input, each divergence is minimized and then manually inspected together with the respective compiler developers, who ultimately confirm whether it is a real bug.

    (3) \textbf{Printer and Runner:} \toolName{} employs three distinct language printers to transform the abstract IR into concrete, interacting source code across different target languages: \texttt{KtIrClassPrinter} for Kotlin, \texttt{JavaIrClassPrinter} for Java and Groovy, and \texttt{ScalaIrClassPrinter} for Scala. The runner then invokes the respective compilers in the following modes:
    \begin{itemize}
        \item \texttt{NormalTest}: Compiles the code using a single compiler version to detect crashes or internal errors.
        \item \texttt{DifferentialTest}: Executes cross-version or cross-compiler trials, where behavioral or bytecode mismatches indicate potential bugs.
    \end{itemize}
    
    (4) \textbf{IR Reductor:} When a bug is detected, \toolName{} serializes the generated test program into the unified IR and minimizes it using our IR reducer. The reducer is based on an optimized Delta Debugging Minimization (DDMin) algorithm and employs an active validator to ensure that the IR remains structurally valid after each element removal. After each reduction step, the resulting IR is translated back into source code and executed to determine whether the bug is still reproducible. If the bug persists, the reduced IR is retained for the next iteration; otherwise, the change is rolled back. This process continues until no further reduction is possible without eliminating the bug. The final minimized IR is then translated into source programs and reported to the corresponding compiler development team for confirmation.
    

\subsection{Unified Intermediate Representation (IR)}\label{subsec:ir_design}

To decouple test generation from language-specific syntax, \toolName{} introduces a unified intermediate representation. The type system and structure of our IR are inspired by the Kotlin compiler's backend IR, which models the rich semantic features necessary for multi-language JVM interoperability—such as advanced generics, mixed nullability, and platform-specific types. Rather than invoking the Kotlin compiler during code generation, \toolName{} implements these structural abstractions independently, enabling flexible, multi-target code generation.

\subsubsection{Program Structure}
The IR models a multi-language program as a hierarchical declaration tree. The root \texttt{Program} node acts as a container for multiple \texttt{ClassDeclaration} nodes. Each class maintains explicit metadata, including a target-language tag, an inheritance chain (\texttt{extends}/\texttt{implements}), optional type parameters, and its member functions. For instance, a cross-language boundary where a Scala class extends a parameterized Java interface is captured as a unified declaration tree before code generation.

\subsubsection{Type System Representation}
To comprehensively stress-test modern JVM compiler front-ends, the IR supports five type forms that map directly to internal compiler metadata structures:
\begin{itemize}
    \item \textbf{Simple types:} Unparameterized base classifiers or primitive representations (e.g., \texttt{Int}, \texttt{String}, \texttt{Any}).
    \item \textbf{Parameterized types:} Generic types bound to explicit type arguments (e.g., \texttt{List<Int>}).
    \item \textbf{Nullable types:} Explicitly nullable forms denoted by a suffix modifier (e.g., \texttt{String?}).
    \item \textbf{Platform types:} Types with ambiguous nullability denoted by an exclamation mark (e.g., \texttt{String!}), replicating how the Kotlin compiler treats unannotated Java types at cross-language boundaries.
    \item \textbf{Type parameters:} Generic variables constrained by optional upper bounds (e.g., \texttt{T <: Comparable<T>}).
\end{itemize}

\subsubsection{Traversal Mechanisms}
The pipeline interacts with the IR through two unified traversal components: a \textbf{top-down visitor} for structural metadata collection and verification, and an in-place \textbf{transformer} for subtree replacement. The generator and the reduction validator use the visitor to enforce structural soundness, while the mutator uses the transformer to transform the IR.

\subsection{Mutation Strategies}\label{subsec:mutation_strategies}

\toolName{} invokes its mutation engine to increase program diversity. Each mutation operator targets a specific semantic feature known to be error-prone during cross-language compilation. As summarized in Table~\ref{tbl:mutations}, \toolName{}'s seven mutation operators target four areas of cross-language divergence.

\begin{itemize}
    \item \textbf{Generic Subtyping.} \texttt{mutateGenericArgumentInParent}, \texttt{mutateGenericArgumentInMemberFunctionParameter}, and \texttt{mutateClassTypeParameterUpperBound} mutate type arguments and bounds, forcing compilers to reason about variance, substitution, and type erasure.

     \item \textbf{Nullability.} Operators like \texttt{mutateParameterNullability} and \texttt{mutateClassTypeParameterUpperBoundNullability} toggle nullability on parameters and bounds, exercising cross-language nullability coercion, where missing
    or misinterpreted metadata frequently triggers compiler errors.
    
    \item \textbf{Override Resolution.} \texttt{removeOverrideMemberFunction} reduces an overridden method to an empty stub, exercising cross-language method resolution and linkage rules, where Java, Kotlin, and Scala diverge.

    \item \textbf{Language placement.} \texttt{shuffleLanguage} reassigns class languages while holding structure fixed, re-exposing the same pattern under different language combinations and amplifying the other operators.
   
\end{itemize}

During execution,  operators are selected probabilistically from a weighted distribution. To increase test-case complexity, multiple mutations can be  chained onto a single IR program before it is passed to the printing layer.

\section{Tool Availability and Running the Tool}\label{sec:running}

\textbf{Tool Availability}: \toolName{} is open-source at \url{https://github.com/XYZboom/CrossLangFuzzer}, with an archived snapshot at \url{https://doi.org/10.5281/zenodo.20925432}. A video demonstration is available at \url{https://youtu.be/XBG6dUO0Adk}.


\subsection*{Docker (Quick Run)}

\begin{verbatim}
docker pull xyzboom123/clf:dev
docker run -it xyzboom123/clf:dev
./quick_run.sh
\end{verbatim}

This runs the Kotlin compiler differential testing in a single command. When a bug is found, the runner stops and reports the minimized reproducer in \texttt{out/min/}.

\subsection*{Gradle Build and Run}

Clone the repository and build the project using Gradle. Each JVM compiler is tested through its own Gradle runner. The following example shows how to execute the Kotlin compiler runner:

\begin{verbatim}
# Kotlin compiler (JDK 17 required)
./gradlew :runners:kotlin-runner:run --args="-s" \
  -Dorg.gradle.java.home=/path/to/jdk17
\end{verbatim}

Commands for the Groovy and Scala runners, along with detailed command-line options, are provided in the repository's README. For each detected bug, the corresponding runner generates a bug report under the \texttt{out/min/} directory, including the minimized source files, the serialized IR, and the corresponding compiler error output.


\section{Results}\label{sec:eval}




Table~\ref{tbl:bug-overview} summarizes the 32 bugs uncovered by \toolName{}. All reports were validated by the corresponding compiler teams, confirming the effectiveness of \toolName{} in revealing real cross-language compilation defects. Notably, these bugs were found in the latest actively maintained compiler versions, demonstrating that cross-language inconsistencies remain a significant challenge and that \toolName{} can effectively expose defects missed by existing testing efforts.

Beyond discovery, the reports have proven actionable. Every issue was verified by the respective development team, and several have already been fixed—most notably, the Groovy maintainers resolved all four reported bugs, and one Kotlinc bug has been patched, with the remaining 14 confirmed and awaiting fixes. 



\begin{table}[t]
\centering
\caption{Overview of the bugs discovered by \toolName{}.}
\label{tbl:bug-overview}
\small
\begin{tabularx}{\columnwidth}{@{}l X c l@{}}
\toprule
\textbf{Compiler} & \textbf{Bug IDs} & \textbf{\#} & \textbf{Status}\\
\midrule
Kotlinc & \kt{74109}, \kt{74147}, \kt{74148}, \kt{74151}, \kt{74156},
          \kt{74160}, \kt{74174}, \kt{74188}, \kt{74202}, \kt{74209},
          \kt{74288}, \kt{78819}, \kt{79508}, \kt{80382}, \kt{80387}
        & 15 & 1 fixed, 14 confirmed\\
Groovyc & \groovy{11548}, \groovy{11549}, \groovy{11550}, \groovy{11579}
        & 4 & 4 fixed (100\%)\\
\addlinespace
Scala3c & \scalaThree{22307}, \scalaThree{22308}, \scalaThree{22309},
          \scalaThree{22310}, \scalaThree{22311}, \scalaThree{22312},
          \scalaThree{22717}
        & 7 & confirmed\\
Scala2c & \scalaTwo{13074}, \scalaTwo{13075} & 2 & confirmed\\
Javac   & \jdk{8347330}, \jdk{8352290}, \jdk{8361835}, \jdk{8370716}
        & 4 & confirmed\\
\bottomrule
\end{tabularx}
\end{table}

\section{Summary and Future work}\label{sec:sum}

In summary, we address the challenge of validating cross-language JVM compilation with \toolName{}, which synthesizes cross-language IR and programs, applies seven mutation operators to diversify test cases, and automatically reduces bug-triggering programs. Evaluated on five JVM compilers, \toolName{} uncovered \countAll{} confirmed bugs, demonstrating its effectiveness in exposing cross-language compilation defects.

Because generation and mutation operate on an abstract IR in \toolName{}, the serialized IR provides a natural interface for LLM-assisted fuzzing. For example, an LLM could directly manipulate the serialized IR or guide mutation selection based on observed error logs. We leave such integrations to future work.

\bibliographystyle{ACM-Reference-Format}
\bibliography{ref}

\end{document}